\documentclass[]{aastex631}
\usepackage{comment}

\shorttitle{Manek \& Brummell}
\shortauthors{Manek \& Brummell}
\graphicspath{{./}{figures/}}
\begin{document}

\title{On the Origin of Solar Hemispheric Helicity Rules: Rise of 3D Magnetic Flux Concentrations through a Background Magnetic Field}
\correspondingauthor{Bhishek Manek}
\email{bmanek@ucsc.edu}

\author[0000-0002-2244-5436]{Bhishek Manek}
\affiliation{Laboratory for Atmospheric and Space Physics, University of Colorado, Boulder, CO 80303, USA}
\author[0000-0003-4350-5183]{Nicholas Brummell}
\affiliation{Department of Applied Mathematics, Jack Baskin School of Engineering, University of California Santa Cruz, \\ 1156 High Street, Santa Cruz, California 95064, USA}

\begin{abstract}
Sunspots and active regions observed on the solar surface are widely believed to be manifestations of compact predominantly-toroidal magnetic field structures (``flux tubes") that emerge by magnetic buoyancy from the deeper interior of the Sun.  Much work has examined the evolution of such magnetic structures, typically considering them as idealized isolated magnetic entities and not as more realistic magnetic concentrations in a volume-filling background magnetic field.  Here, we report results that explore the buoyant rise dynamics of magnetic concentrations in a volume-filling field in the full three dimensions. Earlier 2.5D work in this series \citep{Manek:Brummell:Lee:2018,Manek:Brummell:2021,Manek:Pontin:Brummell:2022} established the remarkable fact that the twist orientation of a flux concentration relative to the background field affected it's likelihood to rise and emerge, regardless of whether the buoyant rise took place in the absence or presence of convection. The contrasting dynamics between structures with differing orientations leads to a selection mechanism that reproduces characteristics of the ``solar hemispheric helicity rule(s)" (SHHR) observations strikingly well.  Here, we show that this two-dimensional selection mechanism persists in the face of the added complexity of three-dimensional dynamics. Arching of the magnetic structure in the third dimension, as might be expected in the solar application, is introduced.  The role of tension force leading to this selection mechanism is elucidated and subtle differences that arise due to the three-dimensional geometry are discussed.

\end{abstract}

\section{Introduction} \label{sec:intro}

The characteristics of sunspots in active regions are the original tell-tale signatures that lead to the conclusion that the Sun's magnetic field is likely generated by a solar dynamo.  Any full magnetohydrodynamic theory for the dynamo should be able to explain the origin of the many surprisingly systematic observational features of sunspots and the active regions they occur in.  For example, the origin of pairs of sunspots, their 11-year cycle of latitudinal appearance (the ``Butterfly diagram"), the polarity reversal of the leading and trailing spots in the pairs every 11 years  \citep[Hale's Polarity Law:][]{Hale:etal:1919}, and the systematic tilt of the sunspot pairs  \citep[Joy's Law:][]{Hale:etal:1919} should all be explained by any theory.  

A more recently developed active region law is the ``solar hemispheric helicity rule" (SHHR).  This empirical rule is derived from observational measurements \citep[see e.g.][]{Pevtsov:Canfield:Metcalf:1995} that are generally proxies for the vertical component (normal to the solar surface) of the magnetic helicity or the magnetic current helicity.  The magnetic helicity, $H_m = \int_v {\bf A} \cdot {\bf B}~dV$, where ${\bf A}$ is the vector potential function for the magnetic field ${\bf B}$, is perhaps the more useful mathematical quantity since it is a conserved quantity in ideal magnetohydrodynamics, but clearly is difficult to compute merely from observations of the magnetic field.  The magnetic current helicity, $H_c = \int_v {\bf J} \cdot {\bf B}~dV$, where ${\bf J} = \nabla \times {\bf B}$, is more readily computed, and proxies for the vertical component of this quantity are expected to have the same sign as proxies for $H_m$ \citep{Gosain_Brandenburg_2019}. The normal component of the current helicity is also nicely interpretable as the twist of the magnetic field where it pierces the solar surface.  The SHHR then states that, in general, this twist is left-handed in the Northern hemisphere and right-handed in the Southern.  The rule is a weak one, however, being only true for roughly two-thirds of active regions studied.  Interestingly though, the rule remains true over both halves of the full 22 year activity cycle, despite the polarity reversal of the magnetic field component normal to the surface in the sunspots themselves.  Clearly, this requires that the surface tangential field components reverse too to preserve the handedness.  Such an interesting persistent correlation could provide a window into the underlying fields, and demands an explanation, as for the other surprisingly ordered rules and laws for sunspots and active regions.

Current global dynamo models \citep[see e.g.][]{Sacha_Miesch_Toomre_ASH_Dynamo_2004, Ghizaru_EULAG_2010, Guerrero_EULAG_Dynamo_2016} quite consistently produce dynamos with cyclic large scale magnetic fields in certain (unfortunately not completely realistic) parameter regimes.  The cycling large-scale toroidal field can be interpreted as evidence of activity possibly related to the origin of a solar-like Butterfly diagram, for example, but the models do not produce magnetic field structures that as yet can be consistently identified as the equivalents of active regions and sunspots.  The examination of active region or sunspot dynamics has therefore become a separate endeavour.  Hopefully these thrusts will be joined consistently at some point in the future, and the theory outlined in this paper is a step in that direction.

Sunspot or active region theory, and simulations designed to examine such phenomena, generally examine compact magnetic structures in isolation from any other large scale dynamo field components.  These model structures, often referred to as ``flux tubes", consist mostly of strong toroidal magnetic flux (perhaps somewhat twisted and writhed to contain a poloidal component) and emerge at the surface, having been formed deeper in the solar interior and subsequently transported upwards by magnetic buoyancy \citep{Parker:1955a}.  The exact point of origin of these and their relationship to the large-scale dynamo field remains uncertain, as mentioned above.  However, the solar tachocline, a region of strong differential rotation at the junction of the convection zone to the deeper radiative zone, is often cited \citep[see e.g.][]{Weiss_1994, Hughes_Rosner_Weiss_2007} as a likely region for the production of such structures, due to the availability of strong velocity shear there to drive the induction of the significant magnetic gradients necessary for efficient magnetic buoyancy.  

There are a number of models that use the ``isolated flux tube" formalism in various guises to explain the origin of the SHHR.  Perhaps the most cited theory to date is that of \cite{Longcope:Fisher:Pevtsov:1998}, where the systematic magnetic helicity of the magnetic structure is derived from the systematic kinetic helicity inherent in the convection that the magnetic structure travels through, with the kinetic helicity bias owing its existence to the background rotation of the star. The connection between these two helicities, kinetic and magnetic, is known as the $\Sigma$-effect, first modelled in the paper of  \cite{Longcope:Fisher:Pevtsov:1998}.  That paper describes an elegant, but highly simplified, mathematical model of the dynamics.  First of all, the model utilizes thin flux tube theory \citep{Spruit:1974} to describe the motion of the magnetic structure where the actual finite size compact magnetic structure is replaced by a single magnetic field line describing its central axis.   This line is then ascribed certain dynamical properties, such as buoyancy and drag, to define its evolution.  In the $\Sigma$-effect theory, an additional property, that of twist, is also assigned and is governed by a separate dynamic equation, as developed by \cite{Longcope:Klapper:1997}.  Here, the evolution of the twist depends on the stretching of the line, the rotation of the line, and a $\Sigma$ term that represents the buffeting of the line by external fluid velocities.  The key development in the \cite{Longcope:Fisher:Pevtsov:1998} paper is that, given a turbulence model for the velocity field that has significant prescribed net kinetic helicity, a systematic writhe of the central line that represents the location  of the thin flux tube should be induced.  If the total helicity is to be conserved, as it should be in non-diffusive dynamics, then a compensating twist in the assigned twist characteristic should also be induced via the $\Sigma$ term.  Interestingly, given the handedness of the solar rotation, the handedness of the twist induced would have the correct hemispherical sign for the SHHR.  Furthermore, the paper finds that, for their chosen turbulence model, significant amounts of twist could be generated this way, something that had eluded other proposed theories.

This very insightful mathematical model is, of course, vastly oversimplified.  For example, the thin flux tube approximation definitely does not follow the real dynamical buoyant rise of finite size compact magnetic structures, which tends to be dominated by the wakes induced by their finite cross section and which depends critically on the twist distribution for coherent rise \citep[see e.g.][]{Emonet:Moreno-Insertis:1998}.  To date, there is no published verification of an appreciable $\Sigma$-effect in more realistic flux tube simulations \citep[and see][for perhaps the opposite]{Wade:2023}.

An alternate theory was introduced and expanded upon in a series of papers by Manek et al \citep{Manek:Brummell:Lee:2018,Manek:Brummell:2021,Manek:Pontin:Brummell:2022}.  This theory is based on a completely different effect and has been validated and demonstrated using simulations of finite-sized magnetic structures under a range of circumstances.
These papers were born out of the idea that the so-called flux tubes are not magnetically-isolated structures but are instead strong magnetic concentrations in an otherwise weaker space-filling field.  The dynamics of isolated and non-isolated structures can be quite different.  The original paper in this series \cite{Manek:Brummell:Lee:2018} found that the alignment and relative strengths of the flux tube's twist field with any horizontal background field influenced the likelihood of buoyant rise.  Overall, there is a significant range of twist strengths (relative to the background field strength) where tubes of one handedness of the twist rise preferentially over tubes with the opposite handedness.  Again, interestingly, when the directionality of the background large-scale poloidal and toroidal fields are taken from the solar context, the handedness preferred by this selection mechanism concurs with the SHHR \citep{Manek:Brummell:2021}.

This mechanism has been shown to be robust, not only in parameter space when the magnetic structure is rising through a quiescent hydrodynamic background containing a prescribed large-scale horizontal magnetic field \citep{Manek:Brummell:2021}, but also when rising through a convective background containing a self-consistent large-scale horizontal background field \citep{Manek:Pontin:Brummell:2022}.  These papers also offer reasonable explanations for other observed nuances of the SHHR, such as the violations leading to the weakness of the rule, and potential solar cycle modulations of the effect.  However, these previous simulations were all performed in 2.5D (where the domain is two dimensional, but all three components of vector fields are considered).  The current paper here extends this work to a fully three dimensional model, where an arched finite-sized three dimensional compact magnetic concentration rises through a volume-filling background unidirectional horizontal magnetic field.  This paper only examines the rise through a quiescent, non-convecting background, but is designed to demonstrate that the added complexity of three dimensional motion does not significantly affect the previously discovered dynamics.

In Section \ref{sec:model}, we describe the three-dimensional numerical model and its setup.  In Section \ref{sec:results}, we present evidence for the selective rise theorized earlier found in the three-dimensional simulations, and describe in detail the causes of the observed dynamics. In Section \ref{sec:discussion}, we discuss the implications of this work and ideas for future work.

\section{Theoretical Model} \label{sec:model}
\begin{figure}[h!]
    \centering
    \includegraphics[width=6cm, height=7cm]{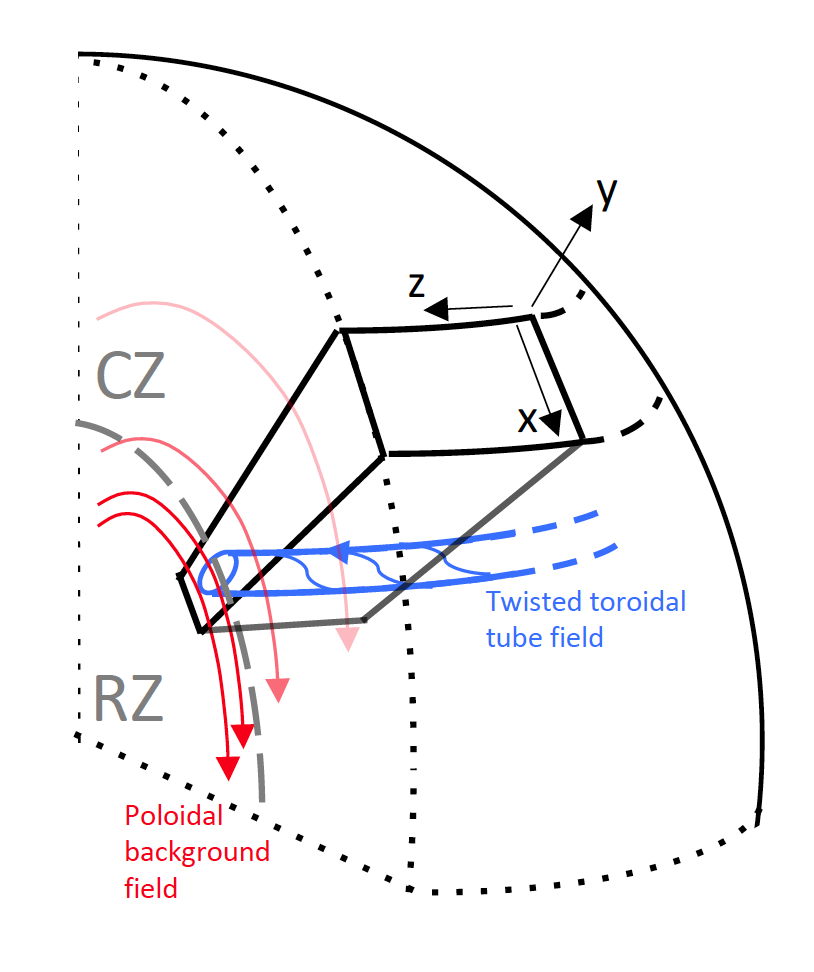}
    \caption{Cartoon sketch of the relationship between our three dimensional Cartesian model and the spherical geometry of a star like the Sun. The Cartesian directions \textit{x} relate to the latitude, \textit{y} to the radial height, and \textit{z} to the longitude. }
    \label{fig:initialsetup}
\end{figure}

The model setup used in this work is the three dimensional extension of \cite{Manek:Brummell:Lee:2018,Manek:Brummell:2021}.   Note that here, for an initial foray into three dimensions, we do not include a convection zone atop a convectively-stable region, as in \cite{Manek:Pontin:Brummell:2022}, but rather consider a single, adiabatic, initially-quiescent layer.  Within this non-convecting layer of fluid, we add magnetic field, in the form of a prescribed cylindrical twisted flux tube embedded deep in a large-scale horizontal background field that varies exponentially in the vertical. The intention of this setup is to simulate (very roughly) the upper tachocline and lower convection zone of a solar-like star.  Figure \ref{fig:initialsetup} shows the relationship between our simple Cartesian simulation domain and the actual spherical  geometry of a solar-like star. The fluid layer is initially quiescent and adiabatic thereby representing the background state of a well-mixed convection and overshoot zone with the actual turbulence ignored.  The flux tube is dominated by its axial component, considered to be originating (most likely) from instabilities of the large-scale dynamo toroidal fields in the deep interior of the star.  Either motions or the origination process also generate a poloidal field component thereby giving the flux tube an initial magnetic twist. The horizontal background field is intended to represent the large-scale dynamo poloidal field, with the (exponentially-varying) concentration towards the lower boundary of the domain a result of convective turbulent pumping \citep[see e.g.][]{Tobias:Brummell:Toomre:2001} of the field out of the convection zone into the overshoot layer below, down the gradient of the turbulent intensity (referred to as the $\gamma$-effect in mean-field theory).  The more self-consistent convective simulations of \cite{Manek:Pontin:Brummell:2022}, that included penetrative convection and the turbulent pumping process (in 2.5D), confirmed that this approximation to the missing dynamics in the earlier works was reasonable and did not affect the major results significantly.  Our initial conditions are then analytical prescriptions for the end state of the following complex system of processes: first, turbulent magnetic pumping drives some fraction of the large-scale poloidal dynamo field from the convection zone into the overshoot zone; then, an $\Omega$-effect mechanism related to inductive stretching of that poloidal field by the differential rotation that is present creates strong toroidal field in the overshoot zone; finally, some instability mechanism (likely magnetic buoyancy)  \citep[see e.g.][]{Vasil:Brummell:2008}  creates the smaller scale twisted magnetic flux concentration embedded in the original large scale background poloidal field.

Our model then solves the equations of fully compressible resistive viscous magnetohydrodynamics using the FLASH code \citep{Dubey:FLASH:2013, Fryxell:etal:2000} in a three-dimensional Cartesian setup. The domain, $ x \in [-1,1] $, $ y \in [0,4] $, $z \in [-2,2] $ has a resolution of $200 \times 400 \times 400$ grid points. We solve the standard equations including all diffusive processes (that is, including the fluid viscosity, thermal conductivity and magnetic resistivity) for the velocity $\textbf{u}=(u_x, u_y, u_z)$, magnetic field $\textbf{B}=(B_x, B_y, B_z)$  and thermodynamic quantities (in particular, temperature, $T$)  \citep[see ][]{Dubey:FLASH:2013, Fryxell:etal:2000}.

The initial conditions pertaining to magnetic fields and the thermodynamic background are prescribed analytically. The magnetic fields in the flux tube are described by $B_{x} = -2q(y-y_{c}), B_{y} = 2qx$, and $B_{z}=1$, within a local radius $r = \sqrt{x^2+(y-y_c)^2}< r_t$ around the center of the flux tube $(x,y)=(0,y_c)$.  The strength of the twist of the field in the tube is given by $q$. This prescription is adapted from the $\alpha = 0$ case discussed in \citep{Hughes:Falle:Joarder:1998}, hereafter referred to as HFJ.  The horizontal large-scale background field is given by 
\begin{equation}
\textbf{B}_{back} (y) = \big( B_{s} \exp \Big(\frac{y_{c}-y}{2H_{b}} \Big),0,0\big),
\end{equation}
where $H_b$ is the scale-height of the field, and $B_s$ represents the strength of the background field relative to the initial axial magnetic field. Importantly, the alignment of the background field in the x-direction depends on the sign of $B_s$. For a positive $B_s$ and a positively twisted ($q > 0$; anticlockwise) flux tube, the background field is aligned with the azimuthal field of the tube at its lower (trailing) edge, and the two fields are opposed at the tube's upper (leading) edge. Note that we interchangeably quote the background field strength in absolute terms (e.g. $B_s=0.1$) or as a percentage of the axial field of the tube (e.g. $B_s=10\%$).

The initial background fluid stratification is defined using a simple polytropic model where the temperature, density, and pressure are given by
\begin{equation}
T = 1 + \theta y'; \hspace{0.1 cm} \rho = (1+\theta y')^{m};
\hspace{0.1 cm} p = (1+\theta y')^{m+1},
\end{equation}
where $y'=4-y$ is the depth, $\theta$ is the imposed temperature gradient, and $m$ is the polytropic index.  Here, thermodynamic variables have been non-dimensionalized by their values at the upper boundary.  The insertion of the magnetic initial conditions into the background stratification can lead to an adjustment of the thermodynamics to accommodate the addition of the magnetic pressure (including that arising from the introduction of the background field) into the total pressure (gas plus magnetic).  It is generally assumed that the total pressure equilibrates quickly.  The accommodation could be achieved by changes to any combination of the temperature and density.  We adopt the method of HFJ and insist that the temperature is continuous horizontally at the edge of the tube but varying inside such that density and total pressure are merely a function of height \citep[see Case 2 in Appendix A of][with Fig. 16(ii) showing plots of these specific initial configurations]{Manek:Brummell:2021}. 
%(see Fig. \ref{fig:initialsetup}e-g).  
This setup leads to a plasma $\beta$ (the ratio of the gas pressure to the magnetic pressure) of $O(10)$, as in HFJ, which is likely lower than expected solar values but is nonetheless dominated by the gas pressure.

We use fixed temperature, stress-free boundaries at the top and bottom, and all other boundary conditions are of outflow type, where the normal derivative is zero.

The setup here is essentially identical to that of \cite{Manek:Brummell:Lee:2018, Manek:Brummell:2021} so far.  A major difference in this three dimensional extension is, however, that we impose a perturbation on the density profile of the flux tube in the axial ($z$) direction such that the central portion of the flux tube is more buoyant compared to the tube at the edges of the domain. That is, the density inside the tube is given by
\begin{equation}
    \rho_{tube}(x,y) = \rho_{init} - \rho' \Big( \text{exp}(-\frac{z^{2}}{2\sigma^{2}}) \Big) + \rho_{adj}, \quad r<r_t
\end{equation}
Here, $\rho_{tube}$ is the final density inside the tube, $\rho_{init}$ is the original density inside the tube 
(previously adjusted as described earlier for the presence of both the tube and background magnetic field components), $\rho'$ is the amplitude of the desired perturbation, $\sigma$ is the lengthscale of the perturbation, and $\rho_{adj}$ is chosen such that the tube ``legs" at the edges of the domain are initially almost neutrally buoyant. 
%\textcolor{blue}{The role of $\rho'$ is to impart a density perturbation to the central portion of the flux tube relative to its legs, and hence it is important to note that $\rho'$ will be different for cases with different $B_s$. }  
We use $\rho' = 0.1 ~\text{max}(\rho_{init})$ and $\sigma=0.375$ and the same value of $\rho_{adj}$ ($=\rho'$) for all the simulations presented in the work. This density perturbation is included with the expectation that the rising magnetic structure may form an arched geometry, reminiscent of the $\Omega$-loop type geometry that is thought to explain the emergence of such magnetic structures at the solar surface as bipolar pairs.  We artificially enforce this here since we are not examining the origin of the structure via magnetic buoyancy, where long axial wavelengths are the preferred instability mode and the geometry therefore would result naturally \citep[see e.g.][]{Newcomb:1961, Matthews:Hughes:Proctor:1995}. Note, that the initial magnetic buoyancy of the tube does depend on the strength of the background field, but only very slightly, being strongly dominated by the tube twist field, which is kept constant in this simulation set \citep[see detailed examination in][in particular, equation (19)]{Manek:Brummell:2021}.

\section{Results} \label{sec:results}

We have run a series of simulations of the model described above for a completely fixed initial configuration except that we vary the strength and orientation of the background magnetic field (relative to the field of the embedded flux tube)\footnote{Note that the theoretical ideas are based on the relative orientation of the background field and the twist, so we could have equally well maintained the direction of the background field and reversed the handedness (changed the sign) of the twist, and achieved the same results.}.
The magnetic flux tube is of radius $r_t=0.125$ centered at $(x_c,y_c)=(0,0.5)$ with axial field $B_z=1$ and a positively-twisted, non-axial (locally-azimuthal) field determined by a twist strength fixed at $q=2.5$ (sufficiently strong that we expect a coherent rise yet sufficiently weak that we do not expect a kink instability).  The large-scale (horizontally-constant) background field has a vertical scale height of $H_b=r_t=0.125$. The only remaining magnetic variables are then the strength of the background field relative to the axial tube strength, $B_s$, and its orientation (determined by the sign of $B_s$).   In this paper, we highlight cases in the range $-0.06 \leq B_s \leq +0.20$ (i.e.~$-6\% \leq B_s \leq +20\%$ of the axial field strength).
The  stratification is weak and adiabatic, specified by $\theta=1.0$ and $m=1.5$.  The (non-dimensional) diffusivities (magnetic, viscous, thermal) governing the equations are $\eta = \mu = 5.e-5$ and $\sigma = 5.e-4$.  

\subsection{Rise Asymmetry in the Presence of a Background Field} \label{sec:main_vol_rendering}
Figures \ref{fig:C2_BGP0_B2_Evolution}, \ref{fig:C2_BGP5_B2_Evolution} and \ref{fig:C2_BGN5_B2_Evolution} show the main results of this paper. Each of these composite plots shows a time sequence covering six different times in the evolution of the magnetic field in a particular simulation via volume renderings of normalized $B^{2}$ ($=B_{x}^{2}+B_{y}^{2}+B_{z}^{2}$).  The three figures show three cases at different background field strengths and orientations, $B_{s}=0.0,~0.05,-0.05$, respectively. 

The first composite sequence, Figure \ref{fig:C2_BGP0_B2_Evolution}, shows the rise of a three-dimensional, twisted flux tube in the absence of a background field ($B_s=0$), i.e., an isolated flux tube. The initial conditions are not in equilibrium state and therefore the flux tube rises at first due to the existence of magnetic buoyancy.  The  circularly-cylindrical tube quickly forms an arched geometry, owing to the imposed initial density perturbation.  The structure's cross-section also quickly becomes distorted into a somewhat squashed ``head'' accompanied by trailing vortices, delineated in the Figures by filaments of the magnetic field that are entrained (see Figure \ref{fig:C2_BGP0_B2_Evolution}b).  The later time dynamics are dominated by the existence of these vortices, with self-advection by this wake taking over the driving of the rise seen in Figures \ref{fig:C2_BGP0_B2_Evolution}c, d, and e. Without the axial density perturbation, we have found that this rise remains independent of the $z$ direction, essentially behaving in the same manner as the earlier 2.5D simulations \citep{Manek:Brummell:Lee:2018, Manek:Brummell:2021}.  The density perturbation introduces some minor three dimensional dynamics other than the arching due to gradients induced along the structure, but nothing major. Our simulation ends when the  central more-strongly buoyant arched region reaches the top of the simulation box (see Figure \ref{fig:C2_BGP0_B2_Evolution}f).  We denote this as a ``successful rise''.  We treat this case as our canonical case for comparison with the others following.

We now focus on the effects of a large-scale background field on these dynamics. Figure \ref{fig:C2_BGP5_B2_Evolution} shows results from the case with $B_{s}=0.05$ where the large-scale background field is moderate in amplitude and oriented in the positive $x$ direction.  Note that the twist of the flux tube is $q=+2.5$, an anticlockwise twist. The initial state can be seen in Figure \ref{fig:C2_BGP5_B2_Evolution}a, which looks similar to Figure \ref{fig:C2_BGP0_B2_Evolution}a except that a layer of $B^2$ is visible at the bottom of the box where the newly-incorporated background field is strongest. However, note that the background field is still weak everywhere compared to the tube. As the flux tube starts rising due to magnetic buoyancy, the connectivity of the large-scale background field overlying the tube impacts the dynamics.  Tension induced in lifting up the overlying field and the wrap of the overlying field around the tube interfering with the vortex wake (compare e.g. Fig.  \ref{fig:C2_BGP0_B2_Evolution}b with Fig. \ref{fig:C2_BGP5_B2_Evolution}c) both act against the rise, but only in a very minor manner at these low background field strengths.  In this case, with the given strength and twist, the initial buoyancy of the flux tube is sufficient to overcome such effects and the rising flux tube diffusively ``detaches'' itself from the influence of the background field, forms vortices, and rises much as in the canonical case (Figs. \ref{fig:C2_BGP5_B2_Evolution}d, e and f). The subsequent rise is qualitatively indistinguishable from the canonical case, yet it should be noted that the flux tube actually takes slightly less time ($t=53$) to reach the top of the simulation domain as compared to the canonical one ($t=59$) due to tension effects in the interior of the tube described in a later section. We again classify this as (at least visually) as a successful rise.

For the next case, we reverse the orientation of the  background field but retain the same amplitude of the field, i.e. we use $B_{s}=-0.05$.  The background horizontal field is now aligned in the negative $x$ direction. Again, all aspects of the flux tube are kept the same, most notably the magnitude and anti-clockwise sense of the twist, $q=+2.5$. Figure \ref{fig:C2_BGN5_B2_Evolution} shows the evolution of the magnetic field amplitude for this scenario. The initial state is shown at time $t=0$ in Figure \ref{fig:C2_BGN5_B2_Evolution}a and can be seen to appear identical to the previous case (since the direction of the background field is not apparent in the variable shown, $B^2$). Remarkably, however, the subsequent evolution is completely different. In this case, the initially buoyant flux tube barely rises, not even in the density-perturbed center.  
Instead of rising as before, the tube essentially remains in place, with only a slight arch due the density perturbation (Figs. \ref{fig:C2_BGN5_B2_Evolution}b,c), and then breaks into two fragments (Figs. \ref{fig:C2_BGN5_B2_Evolution}d) that travel horizontally guided by the background field there (Figs. \ref{fig:C2_BGN5_B2_Evolution}e).
Eventually, a complicated mix of the travelling remnant and the background field attempts to rise due to buoyancy (see Fig. \ref{fig:C2_BGN5_B2_Evolution}f) but nothing succeeds. We classify this as a failed rise.  The dynamics are much more compatible with Alfv\'en wave dynamics on the background field due to a perturbation by the existence of the tube rather than magnetic buoyancy based dynamics.

At this stage, a simple conclusion is that the orientation of the background field has had a very significant impact on the dynamics, at least at these relative field strengths, either allowing or disallowing the rise of a tube.  It is not explicitly demonstrated here but it is true that it is the \textsl{relative} orientation (and strengths) of the background field and the twist of the tube that are the determining factors for this rise selection mechanism; we would have seen exactly the same results if we had fixed the orientation of the background field and simply changed the sign of the twist, $q$, from anti-clockwise (positive) to clockwise (negative).

\subsection{Quantifying the Flux Tube Rise} \label{sec:quantify_ft_rise}

The previous section provided only a qualitative snapshot of the surprisingly selective dynamics in this system. We therefore now establish a more quantitative measure, $y_{ft}(t)$, and examine the fate of the evolution of a flux tube over a broader range of $B_{s}$ strengths (at fixed other parameters). We define $y_{ft}(t)$ as the $y$-coordinate of the maximum axial field ($B_{z}$) of the flux tube in the (axially) central plane ($z=0$) of the simulation box at time $t$. This quantity essentially tracks the vertical location of the peak magnetic field value in the arching loop of the flux tube, and hence is a reliable indicator of whether the tube is rising or not. We curtail $y_{ft}(t)$ when the flux tube either successfully transits roughly through $75\%$ of the height of the simulation box or the trace becomes convoluted enough that we no longer consider it a reliable indicator of position. 

Figure \ref{fig:Track_By_3D_BGP_BGN} shows the measure $y_{ft}(t)$ for simulations where the embedded flux tubes again have the standard positive (anti-clockwise) twist, $q=2.5$, but for a range of background field strengths, $B_s$. Panel $a$ shows positively oriented fields, $B_{s} > 0 $, whereas panel $b$ shows negatively oriented fields, $B_{s} < 0$ (and both show $B_s=0$).  The magnitude of the background field strength varies in a range between $0\%$ and $20\%$ of the axial tube strength. Figure \ref{fig:Track_By_3D_BGP_BGN}a shows that, for positively oriented fields, flux tubes rise successfully from their initial location at $y=0.5$ to $y=3.0$ (75\% of the vertical extent of the domain) when the background field strength is less than $15\%$. The blue triangle markers in Figure \ref{fig:Track_By_3D_BGP_BGN}a correspond to the qualitative result discussed in the previous section (see Fig. \ref{fig:C2_BGP5_B2_Evolution}). On increasing the background field strength further, the rise becomes more difficult and the flux tube structure disintegrates after rising a short distance. In the discrete survey carried out here, it takes a background field strength of $B_{s} > 0.10$ (or $B_s > 10\%$ of the axial tube field strength) before the rise of the tube is strongly impeded. 
On the contrary, when the background field orientation is negative, the rise of the embedded flux tubes are halted for much weaker background field strengths. Figure \ref{fig:Track_By_3D_BGP_BGN}b shows that flux tube structure rises successfully only for very weak background field strength of around $1\%$ or less. The rise is severely impeded for $|B_{s}| \geq 0.03$. The failed rise discussed in previous section for $B_{s}=-0.05$ (as seen in Fig. \ref{fig:C2_BGN5_B2_Evolution}) is (again) shown here with blue triangle markers. 
This quantitative analysis over a broader range of field strengths provides a more comprehensive view of the rise dynamics of the flux tubes in the presence of a background field, but reinforces the previous conclusions that a ``selective rise mechanism" exists, driven by the relative orientation and strength of the background field and the twist of the embedded flux tubes.

Another important detail apparent from this measure, $y_{ft}(t)$, is the impact of the background field on the rise times of the flux tubes. We note that all the successful rise cases in Figure \ref{fig:Track_By_3D_BGP_BGN}a  transit vertically through the domain more quickly than the canonical case (shown in red circles), with the time for the rise monotonically decreasing with increasing $B_s$, until rise actually fails. This suggests that the presence of a positively oriented background field ($B_{s} > 0$) actually facilitates the rise of a  flux tube, whose field is twisted in an anti-clockwise manner, until some critical background field value is reached where different dynamics start to dominate and the rise is halted.  Conversely, when the background field orientation is reversed, the only cases that lead to a reasonable rise are $|B_{s}| \leq 0.01$, (yellow and red diamond markers in Fig. \ref{fig:Track_By_3D_BGP_BGN}b), and the $|B_{s}| = 0.01$ case (yellow markers) is slower to rise than the canonical $B_s=0$ case (red markers). This is an indication that the presence of background field in this orientation actually hinders the rise. 
This behaviour turns out to be a result of the balance between magnetic tensions that are internal and external to the tube as will be explained in the next section.

\subsection{The Role of the Tension Force in the Rise Dynamics}

We have found in the earlier 2.5D simulations \citep{Manek:Brummell:Lee:2018, Manek:Brummell:2021, Manek:Pontin:Brummell:2022} that magnetic tension effects resulting from the combination of the background field with the twist field of the magnetic structure play a crucial role in the main result exhibited here (i.e., that there are differing rise dynamics dependent on the relative orientation of these two components of the field). 
It is not explicitly shown here, but if our simulations are begun without any variance in the third spatial dimension, then the 3D results are qualitatively similar to the 2.5D results of earlier papers, which we will summarise below.  With the introduction of variations in the third spatial dimension, there are possibilities for further tension effects, as will be described.

\begin{figure}
    \centering
    \includegraphics[width=16 cm,height=22cm]{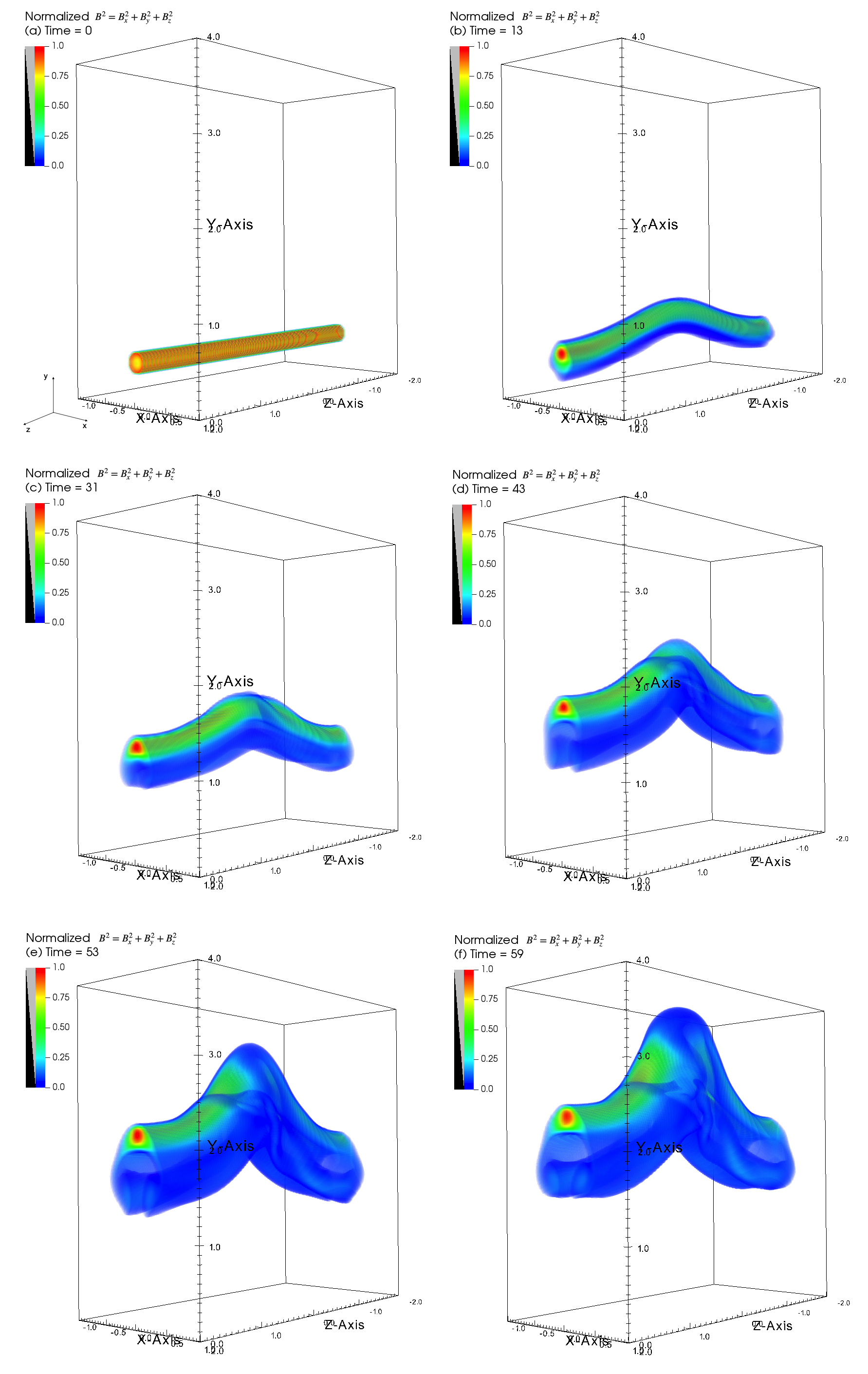}
    \caption{Volume renderings of (normalized) $B^{2}=B_{x}^{2}+B_{y}^{2}+B_{z}^{2}$ for $B_{s}=0$ showing the temporal evolution of a flux tube at six different times. The dynamics at different times shows the arched rise of the tube, with formation of vortices, and the colors show the intensity of normalized $B^2$. We show the orientation of the axes at Time$=0$.}
    \label{fig:C2_BGP0_B2_Evolution}
\end{figure}

\begin{figure}
    \centering
    \includegraphics[width=16 cm,height=22cm]{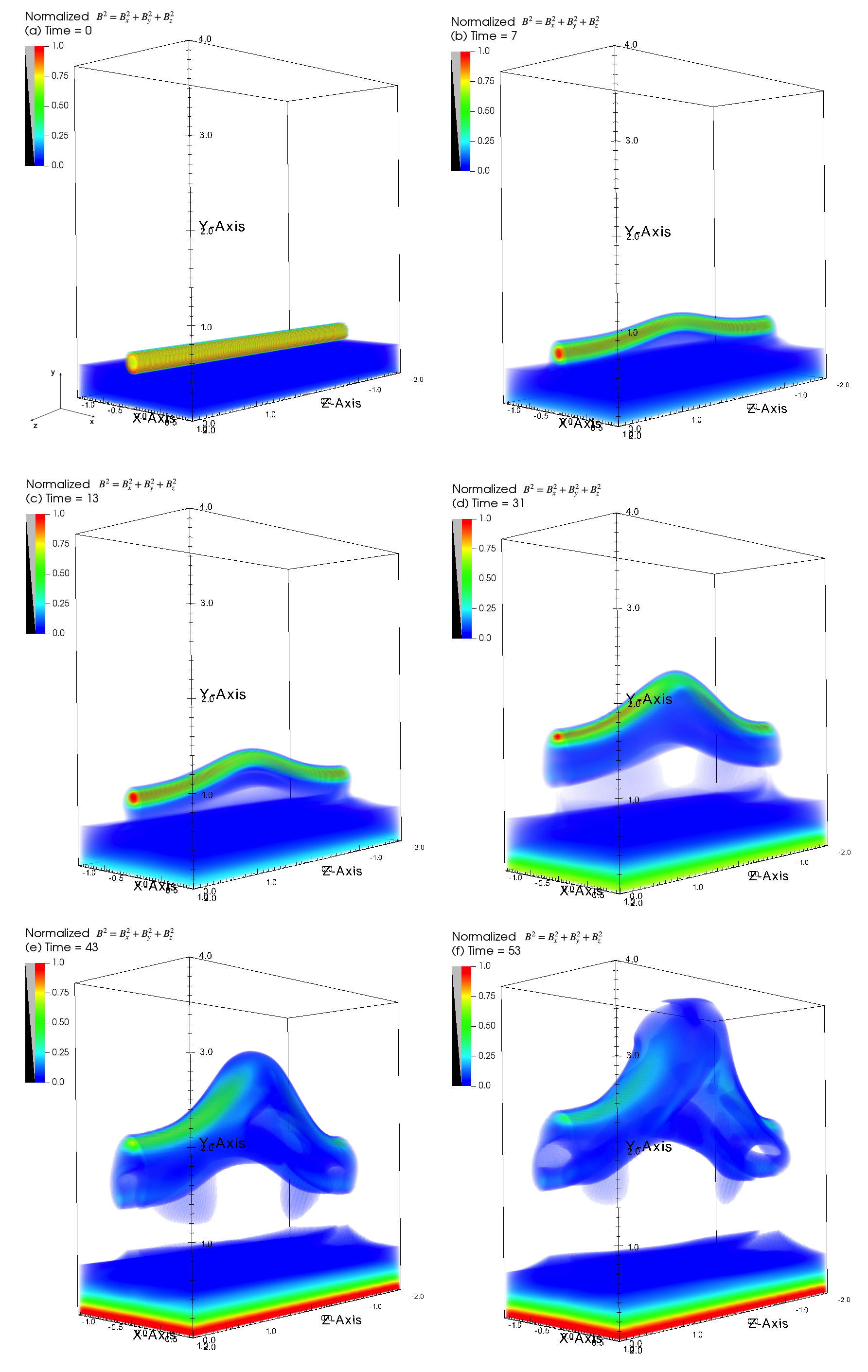}
    \caption{Same as Figure \ref{fig:C2_BGP0_B2_Evolution} but for $B_{s}=0.05$ and at six different times.}
    \label{fig:C2_BGP5_B2_Evolution}
\end{figure}

\begin{figure}
    \centering
    \includegraphics[width=16 cm,height=22cm]{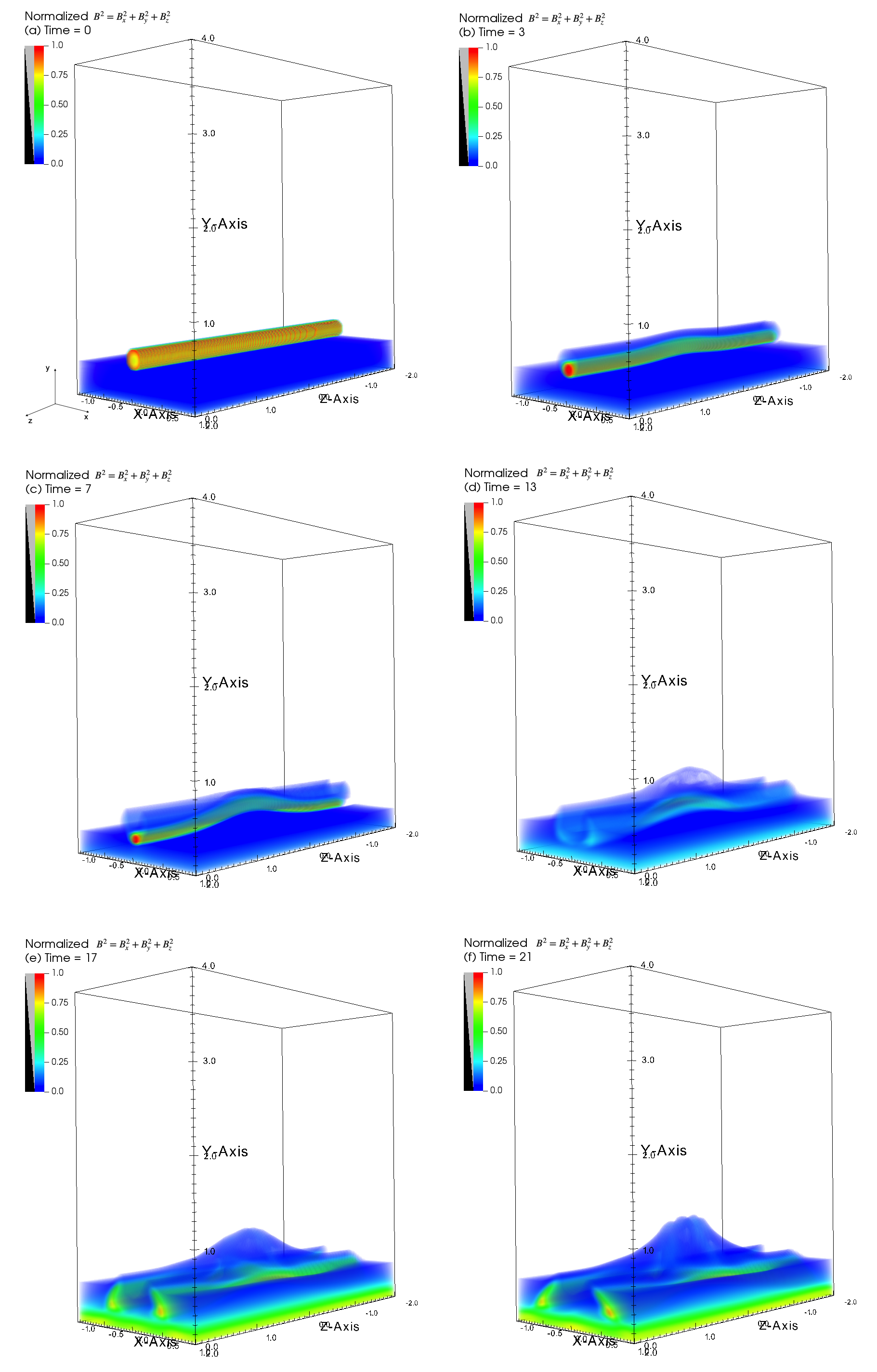}
    \caption{Same as Figure \ref{fig:C2_BGP0_B2_Evolution} but for $B_{s}=-0.05$ and at six different times. }
    \label{fig:C2_BGN5_B2_Evolution}
\end{figure}

\begin{figure}
    \centering
    \includegraphics[width=\textwidth,height=10cm]{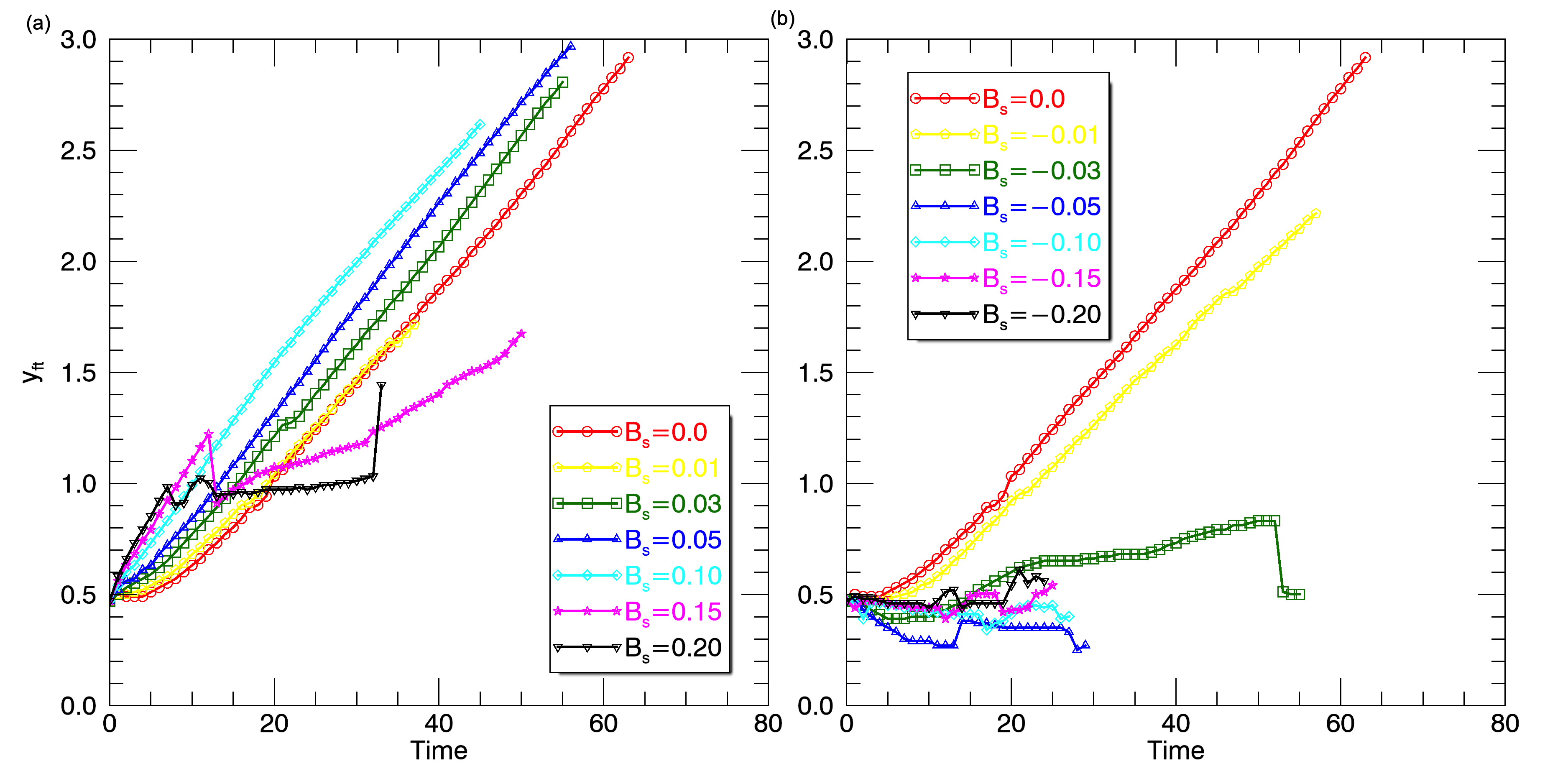}
    \caption{The measure of flux tube location, $y_{ft}$, plotted as a function of time for both (a) $B_{s} \geq 0$ and (b) $B_{s} \leq 0$. Individual cases are curtailed either when the flux tube either transits through $75\%$ of the vertical domain, or fails to rise and/or disintegrates.}
    \label{fig:Track_By_3D_BGP_BGN}
\end{figure}

\begin{figure}
    \centering
    \includegraphics[width=11cm, height=12cm]{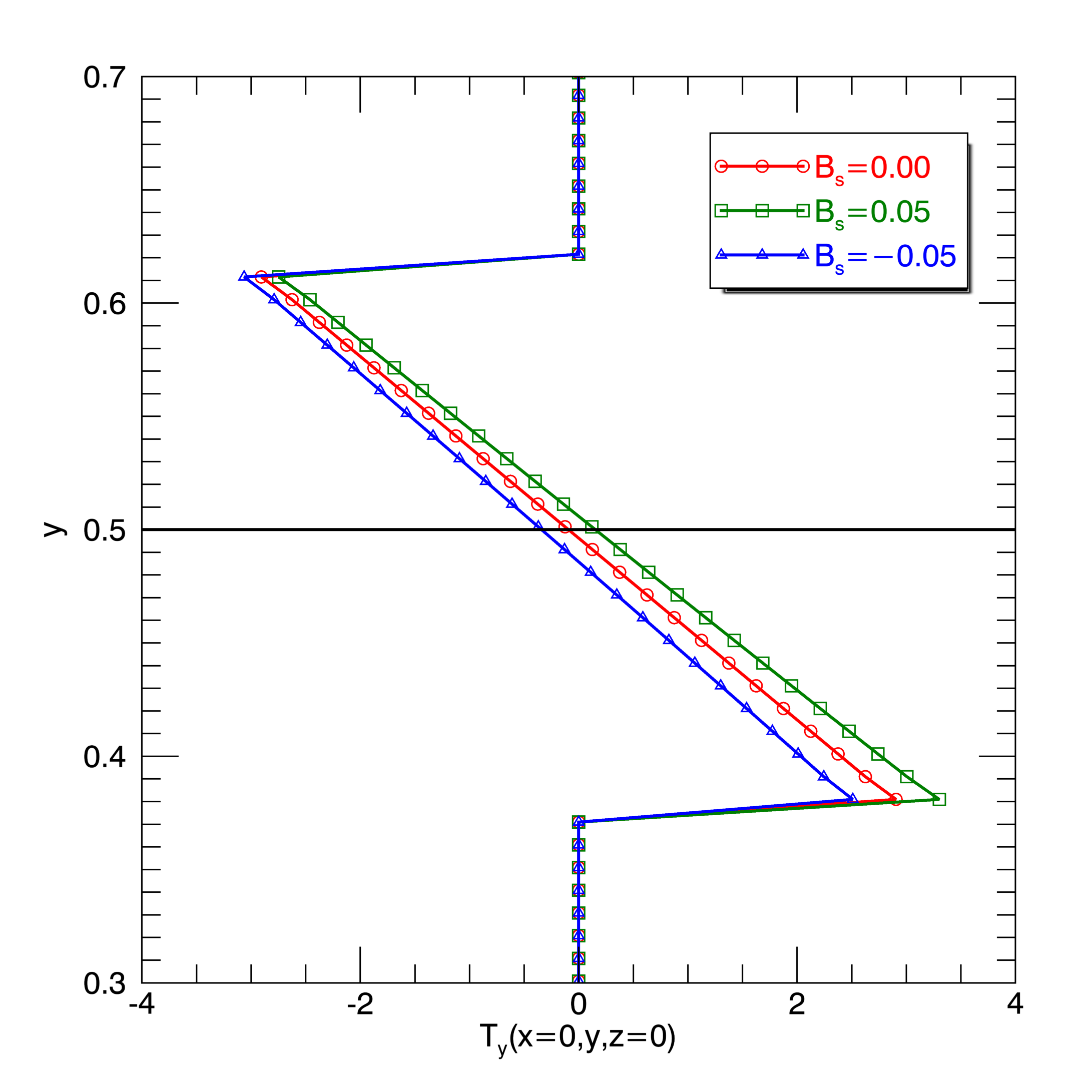}
    \caption{Variation of the vertical tension force plotted as a function of height, $y$, along a linecut at ($x=0,~z=0$) in the initial conditions at $t=0$. Three different background field cases are shown: $B_{s}=0,~0.05,~\text{and}~-0.05$. The figure is zoomed in $y$ to emphasize the dynamics around the location of the flux tube.}
    \label{fig:Tens_y_Linecut_3D}
\end{figure}

In the 2.5D cases, it was shown that the internal tension forces within the magnetic structure engendered by the combination of the azimuthal local field of the structure with the background field in the initial conditions played the dominant role in the differential dynamics.  
To demonstrate this again from the current simulations, Figure \ref{fig:Tens_y_Linecut_3D} shows (on a restricted vertical interval around the flux tube for clarity) profiles of the vertical component of the tension force, $T_y(y) =(B_x\partial_x + B_y\partial_y+B_z\partial_z)B_y$, as a function of height, $y$, over the vertical line-cut at $x=0, z=0$  at the initial time, $t=0$.  The plot shows these  profiles for the three different background field strengths, $B_{s}=0,~0.05,$ and $-0.05$, discussed qualitatively in Section \ref{sec:main_vol_rendering} with corresponding volume renderings in the $t=0$ instances ($a$ panels) of Figures \ref{fig:C2_BGP0_B2_Evolution}, \ref{fig:C2_BGP5_B2_Evolution}, and \ref{fig:C2_BGN5_B2_Evolution}. For the case with no large-scale background field, $B_{s}=0$, this tension force is symmetric about the vertical center of the magnetic flux tube, as shown by the red circles in Figure \ref{fig:Tens_y_Linecut_3D}. This is expected as this tension force is purely due to the uniformly twisted field of the tube, and is inwardly directed towards its center. In this case, this force dynamically only helps to keep the flux tube coherent during rise, and the net upward tension force acting on the tube is zero, leaving magnetic buoyancy as the only magnetic upward force.  In the other two cases, where a large-scale background field has been introduced, the tension force profile becomes asymmetric about the center of the tube, leading to a net vertical force on the tube.   For the positively-directed background field case, $B_{s}=0.05$ (green squares in Figure \ref{fig:Tens_y_Linecut_3D}), the addition of the background field differentially 
(in height) shifts the tension force by a positive value, thereby resulting in a relatively larger upward-directed ($T_y>0$) vertical tension force acting on the lower portion of the flux tube ($0.375 \leq y \lesssim 0.5$)
than the downward directed tension force $T_y<0$ in the upper portion ($0.5 \lesssim y \le 0.625$). 
The overall effect of this is to create a net upwards tension force that assists the magnetically-buoyant rise of the flux tube. 
When the background field orientation is reversed, $B_{s}=-0.05$ (shown with blue triangles in Figure \ref{fig:Tens_y_Linecut_3D}), the opposite happens,  
creating a net downward vertical tension force acting on the flux tube, acting in opposition to the magnetic buoyancy and thus impeding its rise.

Previously, this analysis, as above, was confined to an examination of the initial conditions.
Here, for the first time, we examine whether this effect is the driving force throughout the longer evolution of the magnetic structure and, additionally, examine the extra tension effects that might come into play when the dynamics are not independent of the third spatial dimension.

\begin{figure}
    \centering
    \includegraphics[width=13cm, height=10cm]{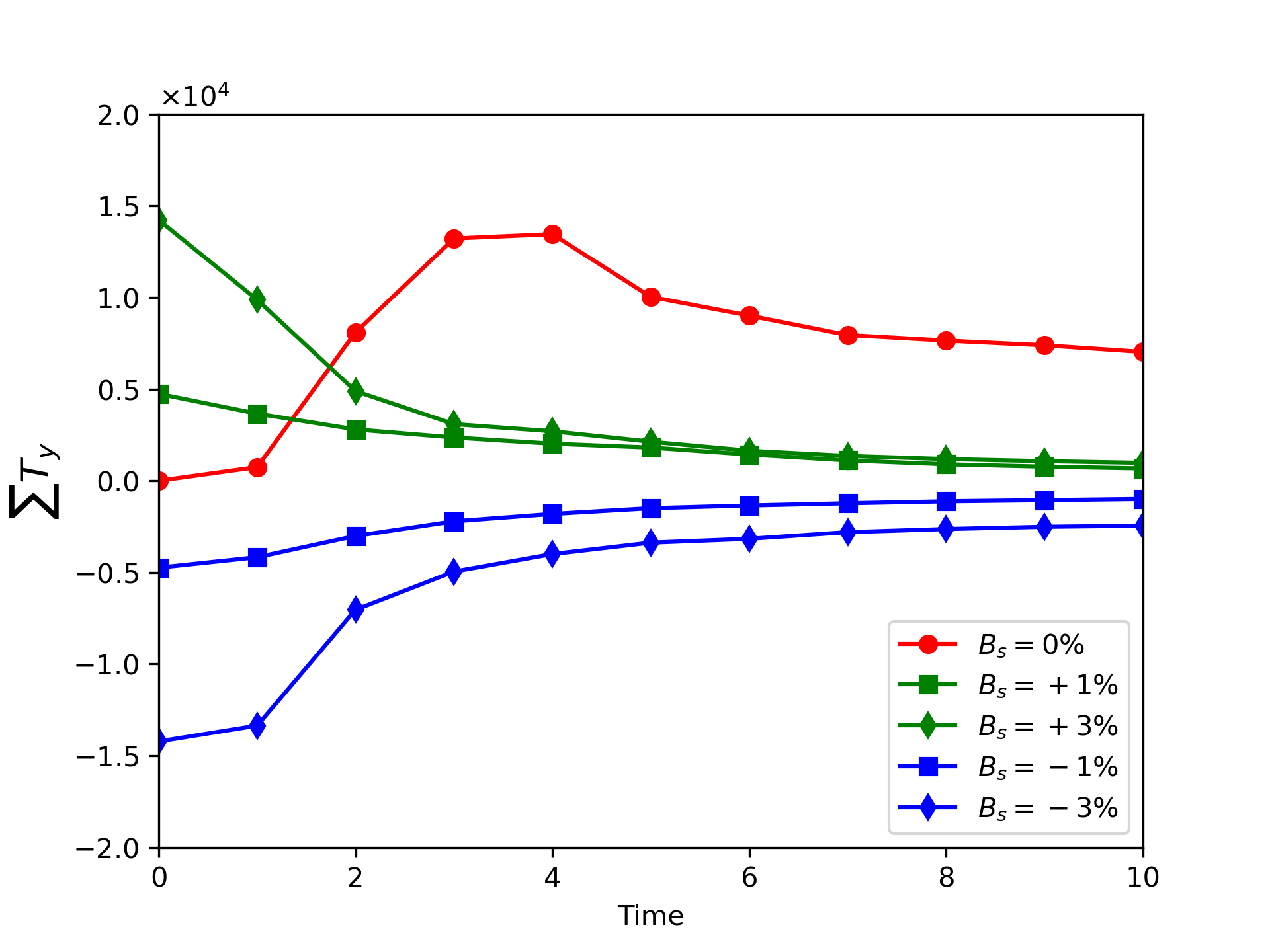}
    \caption{The total vertical tension force integrated over the flux tube as a function of time for different values of the background field, $B_s$.}
    \label{fig:discuss}
\end{figure}

Figure \ref{fig:discuss} shows the total vertical tension force integrated over the volume of the flux tube as a function of time for different background field strengths. In order to ascertain the volumetric location of the flux tube over time, we use a simple mask based on the axial field strength, defining the location of the tube as the set of simulation points where $B_{z}(t) \geq 0.30 ~\text{max}|(B_{z}(t))|$.  The mask value of 0.3 is certainly arbitrary but leads to results that match well with our intuitive visual impression of the location of the magnetic structure (such as those in the volume renderings of Figures \ref{fig:C2_BGP0_B2_Evolution}, \ref{fig:C2_BGP5_B2_Evolution}, and \ref{fig:C2_BGN5_B2_Evolution}) and eliminates contributions from the background field components. 
Figure \ref{fig:discuss} again confirms, this time from a volumetric perspective, that, in the initial conditions ($\text{time,} ~t=0$), the effect of the background field is to either enhance (green cases) or diminish (blue cases) the net vertical tension force experienced by the flux tube (compared to the $B_s=0$ no field case; red circles), depending on the orientation and the strength of the background field. 
However, the figure now also very clearly shows that this initial bias in the vertical tension force in the tube, although diminishing over time, remains significant  for some substantial amount of time (at least $\text{time},~t \leq 10$). 
Even though the tube distorts substantially as it rises, the net tension force remains positive (upward) for positively-oriented background fields and negative for negatively-oriented background fields.  There is a clear separation of the dynamics between the cases associated with the different orientations of the background field for a substantial amount of time.  The fact that this effect diminishes with time as the flux tube rises can be attributed to a combination of expansion of the tube, diffusive dynamics, and drainage of flux out of the tube along the background field.  

The isolated flux tube case, where $B_{s}=0$, shown in red circles in Figure \ref{fig:discuss}, requires some explanation.  Initially, the magnetic structure in this case experiences a net zero volumetric vertical tension force as expected. However, it has an interestingly erratic but net positive value of the volume-integrated vertical tension force as it evolves. As the rise begins, the vertical tension force quickly increases owing to the efficient formation of a distorted tube head coupled with tight trailing vortices.  Eventually, these geometric changes equilibrate, and the behavior smooths out. One of the effects of the background field is to suppress the trailing vortices substantially, and so these effects are not observed in those cases.  It is also interesting to note the significant difference in the value of the total tension force at later times for $B_s=0$, as compared with the other cases. While a little hard to quantify explicitly in a 3D simulation with complicated dynamics, this is likely a reflection of the relative importance of drainage of magnetic flux out of the tube along the background field when it exists overlying the tube (i.e. for $B_s \neq 0$).  In this situation, the flux tube is `leaky', in the sense that no zero magnetic flux surface exists enclosing the magnetic structure, and therefore fluid can exit the magnetic concentration along the complex magnetic field lines connecting it to the background field \citep[as has been expounded in, e.g.,][]{Cattaneo:Cline:Brummell:2006}. With significant drainage, the net magnetic field in the tube contributing to the overall vertical tension force drops, leading to the lower vertical tension forces seen in the figure at later time for these cases. 

Note that in all these cases, there is an additional net upward force owing to the magnetic buoyancy of the tube. While this plays a minimal role in the differential rise of the cases with positively and negatively oriented background field, it is important to account for in the overall rise of the tube (i.e., a negative vertical tension force does not necessarily imply a lack of rise, perhaps just a retardation of the rise). 

\begin{figure}
    \centering
    \includegraphics[width=0.37\textwidth, height=\textheight]{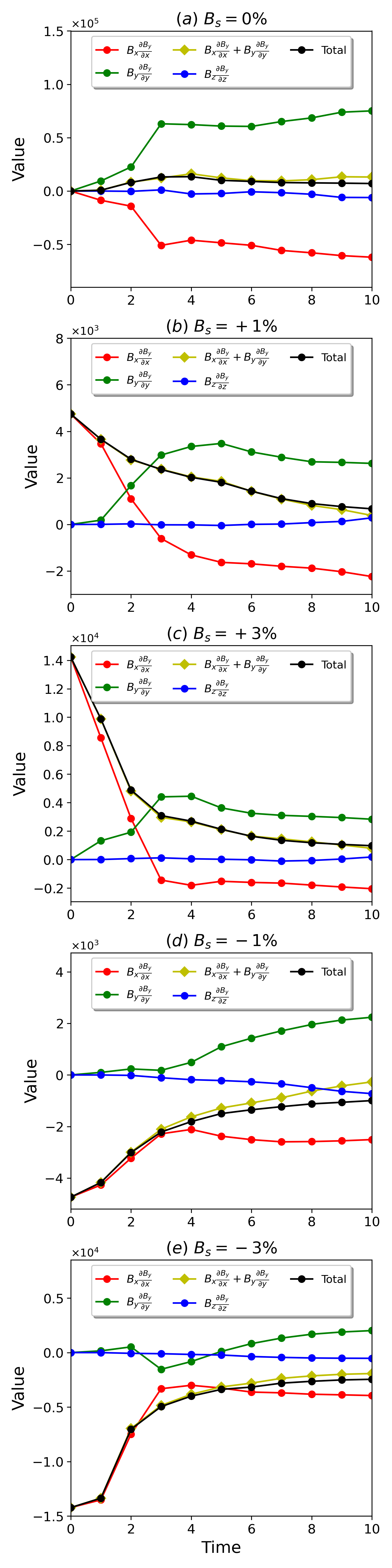}
    \caption{Individual components of the vertical tension force, along with the total tension force, plotted as a function of time for $(a)~B_s=0\%,~(b)~B_s=1\%,~(c)~B_s=3\%,~(d)~B_s=-1\%,~(e)~B_s=-3\%$. All the forces are integrated over the volume of the flux tube.}
    \label{fig:tension_line_all}
\end{figure}

A deeper understanding of the vertical tension forces can be achieved by looking at its constitutive components.  Here, our new 3D cases have more complexity than the older 2.5D cases; the 3D cases allow for tension forces to act both across the tube ($x-y$ plane), related to the overlying background field, and along the tube ($y-z$ plane), due to the arching of the tube in the axial direction.  We show in Figure \ref{fig:tension_line_all} the evolution of the volumetric contributions of the separate components of the vertical tension force ($B_x \partial_x B_y$, $B_y \partial_y B_y$, $B_z \partial_z B_y$, where $\partial_x \equiv \partial/\partial x$, etc) within the magnetic structure (as delineated earlier) for different strengths of the background field.  
The terms $B_x \partial_x B_y$ and $B_y \partial_y B_y$ contribute to the across-tube vertical tension, whereas the terms $B_y \partial_y B_y$ and $B_z \partial_y B_z$ contribute to the along-tube tension.
There are two broad conclusions to draw from all the individual cases shown in Figure \ref{fig:tension_line_all}: (a) the vertical tension force dynamics are dominated by the across-tube components (yellow diamonds), and (b) the purely along-tube component, $B_z \partial_z B_y$ (blue circles), has minimal effect on the overall vertical tension force. In the cases with background field ($B_s \neq 0$), the vertical tension force is dominated by the term $B_x \partial_x B_y$ that originates from the background field in the initial stages of the flux tube evolution. This across-tube component of the vertical tension force is the major source for the dramatic change in dynamics as the background field is introduced to an otherwise isolated flux tube. Note again that this term accounts for the entire difference initially, at $t=0$ (the red lines are coincident with the black for each case).  As the tube evolves further and if it rises, the other across-tube term, what might be called the ``self-tension" term\footnote{Note that this ``self-tension'' term is irrelevant if one examines the complete vertical Lorentz force, since this term cancels identically with a portion of the magnetic pressure.  Our interpretation stands only since we are examining the magnetic tension in isolation.} , $B_y \partial_y B_y$ (green circles), becomes increasingly important until it saturates at some value. This term, arising from the geometrical deformations that the tube experiences during its ascent (both across-tube via expansion and the creation of trailing vortices, and along-tube by stretching during arching), supplements the vertical tension force.  It can be seen in the $B_s=0$ case that a similar effect happens in the $B_x \partial_x B_y$ term (red circles) when there is no background field to dominate this term and only the tube interior field to contribute.  These two terms ($B_y \partial_y B_y$ in green and $B_x \partial_x B_y$ in red) are related to the cross sectional expansion and would cancel exactly in this case but for the asymmetry of the distortions.  In the cases where $B_s \neq 0$, this effect seems to be responsible for the reduction of the influence of $B_x \partial_x B_y$ (red circles) in the total tension with time and the simultaneous increase in $B_y \partial_y B_y$ (green circles).  Note, however, that while the rise of the tube is markedly three dimensional, the contribution of the major along-tube component $B_z \partial_z B_y$ to the total vertical tension force is non-zero but certainly weak compared to all other components. The self-tension component $B_y \partial_y B_y$ also contributes to the along-tube total force, but in the arched geometry, downward tension forces are experienced at the apex of the arched tube, whereas upward tension forces are experienced in the curved legs.  These effects apparently tend to cancel, at least in the simulations run here with the current arching geometry.

The above measures have carefully concentrated on the interior field of the magnetic structure.  Other aspects of the role of tension in the background field carry over from 2.5D to 3D.  The rise of the structure within the background field leads to a downward-directed tension above the tube resulting from the lifting of the overlying field by the rising structure.  This acts against the magnetic buoyancy and slows the rise of the tube compared to a case with no background field.  This lifted field wrapping around the structure also affects the vortex wake that dominates the dynamics at later time, leading to suppressed rise too.   Furthermore, the threading of the tube with the background field can lead to drainage of the interior magnetic flux ($B_z$) that provides the initial buoyancy away from the structure, also leading to loss of rise. It is some combination of these three effects that kill the rise of the tube when the overlying field is sufficiently strong.  All of these effects are examined in detail in the earlier 2.5D papers and carry over the 3D situation here.  

One interesting difference between the 2.5D and the 3D cases is the fact that the threshold values of background field that lead to suppression of flux tube rise are somewhat different between the 2.5D and 3D cases, given the limitation that these studies only examine a discrete range of $B_s$.  In the earlier models, a positive background field of about 16-25\% (relative to the axial field) and a negative background field of about 6\% suppressed the rise of a structure (for the similar values of other parameters, ignoring the three dimensional density perturbation).  Here, those same numbers are 15\% and 3\%.
This is perhaps understandable from the fact that the 3D flux tube, as it rises in an arching fashion, will have magnetic flux `drained' from the arching head to its relatively less buoyant footpoints, although, again, drainage is a hard quantity to quantify or demonstrate practically.  Regardless, with this drainage, the effective strength (defined by the amount of magnetic flux) and thereby the magnetic buoyancy force of the rising central portion of the flux tube is weaker leading to a relatively smaller threshold of background field needed to create enough tension to halt the rise. This, however, does not influence the main conclusions of this work, i.e., that there exists an asymmetry in these cutoff values arising from the relative orientation and strength of the flux tube twist and the background field.

\section{Discussion and Conclusions} \label{sec:discussion}
In this work, we have shown that the buoyant rise dynamics of three dimensional twisted magnetic flux tubes through a large-scale background field are distinctly different from those of isolated magnetic structures. The brief survey of parameters presented in this three dimensional simulation suite (specifically the strength and orientation of the background field) shows that the relative orientation and strength of the overlying background field as compared to the local azimuthal twist magnetic field of the magnetic concentration dictates a bias in the rise dynamics, encouraging the rise of one alignment whilst discouraging the rise of the other. Physically, it is the action of asymmetric vertical tension forces acting within the flux tube due to the presence of the background field that engender the potentially disparate dynamics. When the background field is aligned with the local azimuthal twist field at the bottom of the magnetic structure, vertical tension forces act in accord with magnetic buoyancy, encouraging its rise, whereas the opposite alignment pits vertical tension forces against magnetic buoyancy, discouraging rise.
These results qualitatively reaffirm those presented in our previous series of 2.5 dimensional studies  \citep{Manek:Brummell:Lee:2018, Manek:Brummell:2021, Manek:Pontin:Brummell:2022}.  In the three dimensional simulations presented here, the magnetic structure was forced (via initial conditions) to be fully three dimensional (arching).  The earlier theory was validated and appears robust even with the added complexity arising from the three dimensional nature of the dynamics.   The origins of the bias in terms of the vertical tension forces were investigated in more detail here, showing ultimately that the across-tube components of the vertical tension force arising from the background field are the main dynamical contributions, whilst the along-tube components of the tension forces involving the axial flux tube field are not substantially influential.  In this study, we also showed that the effect of these forces is not solely an initial condition effect, but persists for a significant fraction of the evolution of the flux structure.

While the important and unique feature reported in this particular paper is the three dimensionality of the dynamics (and this was a significant and computationally intense step forward), more work is still needed. A more representative three dimensional modeling of flux tubes in the presence of background field for the solar case would involve the presence of convective dynamics and rotation. The introduction of convection into such three dimensional simulations is challenging in various ways. Firstly, of course, resolving such turbulence requires higher resolution and therefore more intense computations.  Secondly, establishing the vertical profile of the background field self-consistently via turbulent pumping is far more complex in the 3D cases than it was in the 2.5D model in \cite{Manek:Pontin:Brummell:2022}.  Controlling the amplitude of the pumped large-scale field is difficult since the diffusive nature of such high-dimensional simulations can quickly lead to a scenario where the pumped field is relatively too weak to influence the flux tube. 
However, such issues can likely be overcome, and we will proceed down this line in the future.

In the broader context of solar observations, this line of work provides an alternate mechanism to explain the origin of helicity (magnetic and current) that follows specific sign rules at the solar surface.  The origin, transport and reconfiguration of magnetically-based helicity is important to our understanding of space weather, and plays a crucial role in dynamo modeling. The previous leading theory, the $\Sigma$-effect of \cite{Longcope:Fisher:Pevtsov:1998}, argued that the buffeting action of rotating convective turbulence on rising magnetic flux tubes can induce the biased twist that matches the solar observations.  The $\Sigma$-effect is pitched as a mechanism by which a flux tube acquires the correct twist as a transfer of kinetic helicity into magnetic helicity.  Our mechanism is not an acquisition of magnetic twist but is instead a selection mechanism. Our theory assumes that a spectrum of twists are created in the flux tube origination process, and then subsequently the ``correct" helicity flux tubes (matching the observations) emerge because they are more likely to rise due to the selection process.  In our case, the selection is a filter that operates due to the action of the large-scale background field. The filtering does indeed match the SHHR, as explained in earlier papers \citep{Manek:Brummell:Lee:2018, Manek:Brummell:2021}.

Both types of mechanisms (our filter mechanism and the acquisition of magnetic helicity from kinetic) could indeed be operating in the solar context.  Interestingly, these two possibilities as currently envisaged would not be co-spatial in the solar application. Our selection mechanism operates most pertinently at the location of flux tube creation, most likely in the tachocline, whereas the $\Sigma$ mechanism operates subsequently in the transit throughout the convection zone.  Comprehensive studies quantitatively distinguishing the role of $\Sigma$-effect and our selection mechanism are currently lacking.  Preliminary modeling of the deep interior \citep{Wade:2023} and near-surface observational results \citep{Yang:2024}, however, both conclude that there is no clear correlation between kinetic helicity and magnetic helicity, at least in the contexts studied. 
A comprehensive simulation study comparing the two mechanisms is required, but suffers the numerical difficulties mentioned above.  We hope to make progress in this area soon.

\begin{acknowledgments}
This research was supported by the National Science Foundation, under grants NSF AST-1908010, and by the joint NASA-NSF Diversity, Realize, Integrate, Venture, Educate (DRIVE) Science Center (DSC) Phase I grant 80NSSC20K0602 and Phase II grant 80NSSC22M0162 for the Consequences Of Fields and Flows in the Interior and Exterior of the Sun (COFFIES) DSC (subcontract to University of California, Santa Cruz). BM was also supported by NASA HSR grant 80NSSC24K0270, NASA SSW grant 80NSSC19K0026, and NASA HTMS grant 80NSSC20K1280. The authors also acknowledge NSF XSEDE allocations for access to the Texas Advanced Computing Center (TACC \url{http://www.tacc.utexas.edu}) at The University of Texas at Austin for providing HPC, visualization, and database resources that have contributed to the research results reported within this paper. We further acknowledge use of the Lux supercomputer at the University of California Santa Cruz, funded by NSF MRI grant AST 1828315. Volume rendering plots in this work were made using the VisIt visualization software \citep{Visit_Visualization}. We thank the anonymous referee for helpful suggestions.
\end{acknowledgments}

%\appendix
%\section{Appendix information}

\bibliography{References_Bhishek}{}
\bibliographystyle{aasjournal}



\end{document}